\documentclass[%
aps,
prb,
amsmath,amssymb,
preprint,%
superscriptaddress,
]{revtex4-1}

\usepackage{graphicx}
\usepackage{dcolumn}
\usepackage{bm}

\begin{document}


\title{Surface acoustic wave coupled to magnetic resonance on a multiferroic CuB$_2$O$_4$}

\author{R. Sasaki}
\affiliation{ 
Department of Basic Science,  University of Tokyo, Tokyo 153-8902, Japan
}%

\author{Y. Nii}%
\affiliation{
Institute for Materials Research, Tohoku University, Sendai 980-8577, Japan
}%

\author{Y. Onose}
\affiliation{
Institute for Materials Research, Tohoku University, Sendai 980-8577, Japan
}%


\begin{abstract}
We observed surface acoustic wave (SAW) propagation on a multiferroic material CuB$_2$O$_4$ with use of two interdigital  transducers (IDTs). The period of IDT fingers is as short as 1.6 $\mu$m so that the frequency of  SAW is 3 GHz, which is comparable with that of magnetic resonance. In antiferromagnetic phase, the SAW excitation intensity varied with the magnitude and direction of the magnetic field, owing to the dynamical coupling between SAWs and antiferromagnetic resonance of CuB$_2$O$_4$. The microscopic mechanism is discussed based on the symmetrically allowed magentoelastic coupling.
\end{abstract}

\maketitle

%


Multiferroics are materials where magnetism and ferroelectricity coexist. 
 Novel electromagnetic phenomena are frequently observed, thanks to the interplay between the magnetism and ferroelectricity.
For example, they show giant magnetoelectric effects, which  is  polarization change induced by a magnetic field, and reciprocally magnetization change induced by an electric field\cite{Cheong2007,Tokura2014}.
The interplay is valid even for the dynamical state.  The magnetoelectric correlation in optical frequency range gives rise to nonreciprocal directional dichroism\cite{Kubota2004,Fiebig2005,Saito2008,Kezsmarki2011} .  The electric-field active magnon mode has also been observed in multiferroics\cite{Pimenov2006}.
Here, we study the dynamical coupling between an antiferromagnetic magnon and surface acoustic wave (SAW)  in a multiferroic CuB$_2$O$_4$.

The SAW is an elastic wave localized on a surface of media\cite{morgan1985surface}. The amplitude decays exponentially with the depth from the surface. 
The SAW can be excited and detected on a piezoelectric substrate with use of interdigital transducers (IDTs). 
The SAW can carry electromagnetic signals between two separated IDTs when the wavelength coincides with the IDT finger period.
The SAW devices composed of two IDTs on a piezoelectric substrate are industrially used as bandpass filters or delay lines. 
The combination with magnetism seems useful for making these devices more functional.
In fact, introducing ferromagnetic thin film between two IDTs gives rise to emergent functionality such as acoustically driven ferromagnetic resonance\cite{Weiler2011,Dreher2012,Labanowski2016}, acoustic spin pumping\cite{Weiler2012,Xu2018}, and nonreciprocal SAW propagation\cite{Sasaki2017}.
A more direct way of introducing magnetism is replacing piezoelectric substrate with multiferroic one.
Quite recently, we succeeded in fabricating SAW device based on a multiferroic material BiFeO$_3$ in collaboration with other researchers\cite{Ishii2018}.  We observed that the SAW intensity and velocity were modulated due to the static magnetostructural change in the magnetic fields.
One might think the dynamical coupling between SAW and magnon should show more rich phenomena but magnon mode in BiFeO$_3$  is too high to be coupled to SAW\cite{Nagel2013,Caspers2016}.
In order to elucidate this issue, we investigate SAW coupled to magnon mode on another multiferroic material CuB$_2$O$_4$. 

CuB$_2$O$_4$ has the non-centrosymmetric but non-polar crystal structure with space group of $I\bar{4}2d$. 
According to symmetry analysis, the piezoelectric tensor is expressed as
\begin{equation}
\left(
\begin{array}{cccccc}
0&0&0&d_1&0&0\\
0&0&0&0&d_1&0\\
0&0&0&0&0&d_2
\end{array}
\right),
\end{equation}where $d_1$ and $d_2$ are non-vanishing constants.
As discussed later, the SAW can be generated in some choice of the device plane and propagation direction.
At low temperature, CuB$_2$O$_4$ exhibits two successive magnetic phase transition at $T_N$ = 21 K and $T^*$ = 9 K. At $T_N$, it shows easy-plane N\'eel type antiferromagnetic order\cite{Boehm2003}.  Ferroelectric polarization depending on the direction of antiferromagnetic moments is emergent in this magnetic phase. Several unusual optical phenomena have been reported such as magnetic field induced second harmonic generation\cite{Pisarev2004} and magneto-optical dichroism\cite{Saito2008}.
The second transition at 9 K corresponds to the incommensurate helical ordering.
In these magnetic states, a magnetic resonance mode was observed in low frequency range of several GHz\cite{Pankrats2000,Fujita2006,Nii2017}, owing to the small magnetic anisotropy of Cu$^{2+}$ $S=1/2$ moment.
SAW excitation in this frequency range can be achieved by means of conventional electron beam lithography technique.
Therefore, this material is suitable for the investigation of the SAW coupled to the magnetic resonance. 

\begin{figure}
  \centering
  \includegraphics[width = \linewidth]{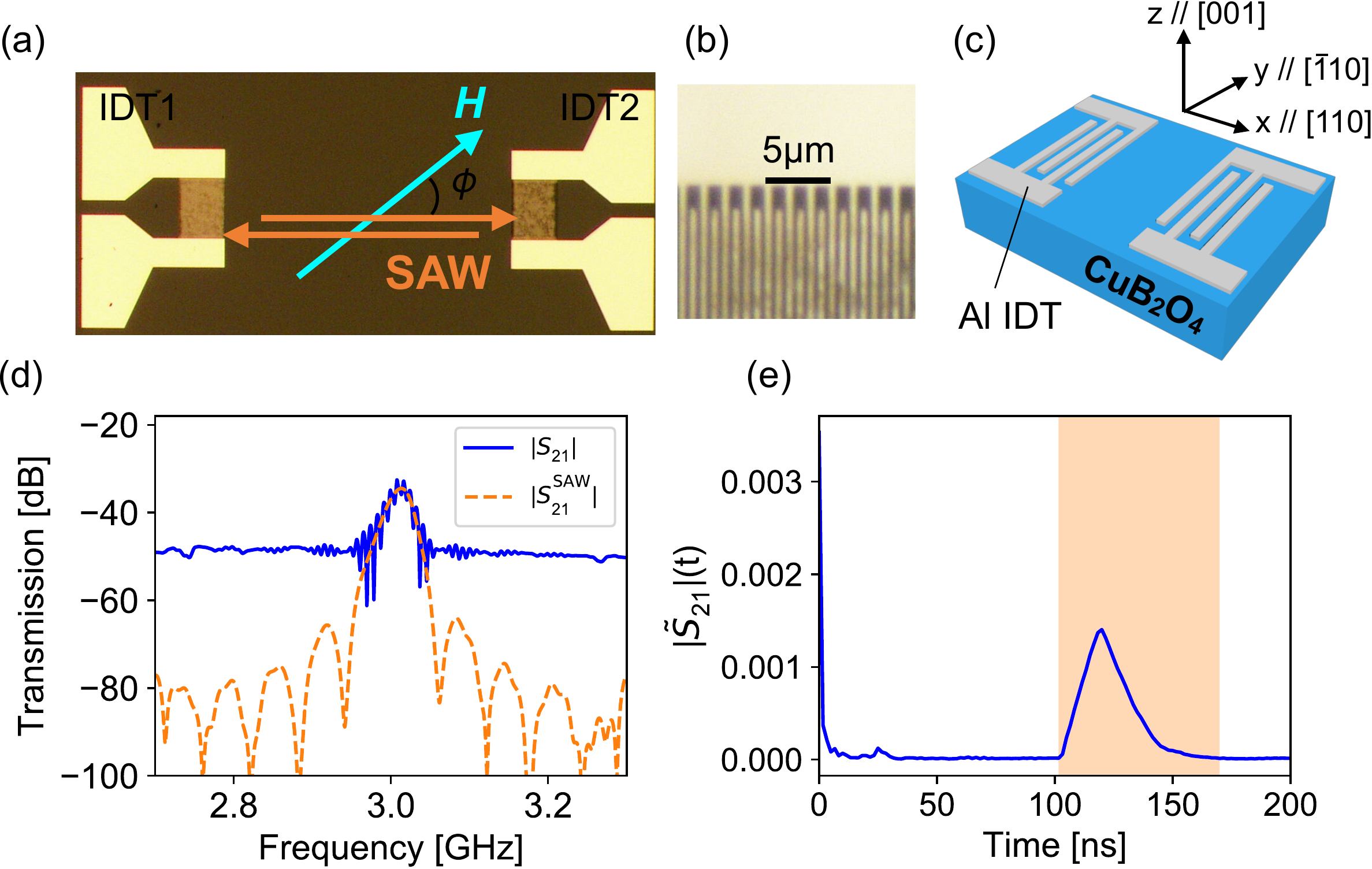}
  \caption{
  (a) Top view of CuB$_2$O$_4$ SAW device used in this research. Magnetic field was applied along the surface of the device. 
  (b) Enlarged view of Al IDT. 
  (c) Schematic diagrams of the device. the surface of device is (001) plane of the CuB$_2$O$_4$ crystal, and SAW propagation direction is parallel to the $\left[110\right]$ axis. 
  (d) Transmission between two IDTs. Solid line is the raw data measured and dashed line shows the transmission due to SAW deduced by the Fourier transformation analysis (see text).
  (e) Transmission as a function of time obtained by the inverse Fourier transformation. Data in the colored region  is used for the Fourier transformation to the SAW transmission $S^{\rm SAW}_{21}$.
  }
  \label{fig:1}
\end{figure}

We fabricated a SAW device on CuB$_2$O$_4$ substrate (Fig. 1(a)).
The CuB$_2$O$_4$ single crystal was synthesized by flux method \cite{Petrakovski2002}. Two Al IDTs with the thickness of 50 nm were fabricated on the substrate using electron beam lithography and electron beam evaporation. One finger width of the IDT and space between the fingers were designed to be 400 nm (Fig. 1(b)) so that the wavelength of SAWs is 1.6 $\mu$m, which is determined by the periodicity of IDTs. The distance between the center of the two IDTs is 580 $\mu$m.
We designed the CuB$_2$O$_4$ SAW device so that the surface of the substrate is perpendicular to crystal $[001]$ axis, and SAW propagation direction is parallel to $[110]$ axis (Fig. 1(c)).
When $x,y,z$ coordinate is defined as $x||[110]$, $y||[\bar{1}10]$, and $z||[001]$, alternating electric field induced  on the IDT has $x$ and $z$ components, and Rayleigh-type SAW has $\epsilon_{xx}, \epsilon_{zz}, \epsilon_{zx}$ components\cite{morgan1985surface}. In this coordinate system,  the piezoelectric tonsor can be discribed as
\begin{equation}
\frac{1}{2}
\left(
\begin{array}{cccccc}
0&0&0&0&2d_1&0\\
0&0&0&-2d_1&0&0\\
d_2&-d_2&0&0&0&0
\end{array}
\right).
\end{equation}
Because $\epsilon_{xx}, \epsilon_{zx}$ components can be induced by the IDT electric field, the SAW can be excited in this configuration.
All the measurements were performed at $T$ = 10 K. As shown in Fig. \ref{fig:1} (a), the magnetic field is applied parallel to the device surface, and the azimuth angle of the magnetic field from the SAW propagation direction is defined as $\phi$.

Figure 1(d) shows the absolute value of complex forward transmission  from left IDT (IDT 1) to right IDT (IDT 2)  $(|S_{21}|(f))$ measured by a vector network analyzer (Agilent E5071C).
We have found a broad peak with small ripples around 3 GHz.
In order to confirm this is the SAW signal, we perform a time domain analysis.
Figure 1(e) shows absolute value of the time domain transmission complex amplitude $\tilde{S}_{21}(t)$ obtained by an inverse Fourier transformation\cite{Kobayashi2017}. 
The large impulse observed around 0 ns is due to the direct electromagnetic transmission between two IDTs. 
In addition, we have observed the delayed impulse around 120 ns. From the distance between IDTs, the velocity of the delayed transmission signal is estimated as $4.8\times 10^3$ m/s, which almost coincides with the phase velocity estimated from the frequency and IDT finger period. Because the velocity is comparable with SAW velocity of other piezoelectrics (e.g. 3488 m/s for YZ-cut LiNbO$_3$\cite{morgan1985surface}), the delayed signal can be attributed to the SAW signal.
We obtained the absolute value of SAW transmission spectra $|S_{21}^{\rm SAW}|(f)$ by performing Fourier transformation only on the SAW transmission time region (the colored region in Fig.  1 (e)).
In this spectrum, the ripples are removed, and the background level is decreased. Hereafter, we used the  $|S_{21}^{\rm SAW}|(f)$ spectra for the analysis of the SAW transmission.

\begin{figure}
  \centering
  \includegraphics[width = 10 cm]{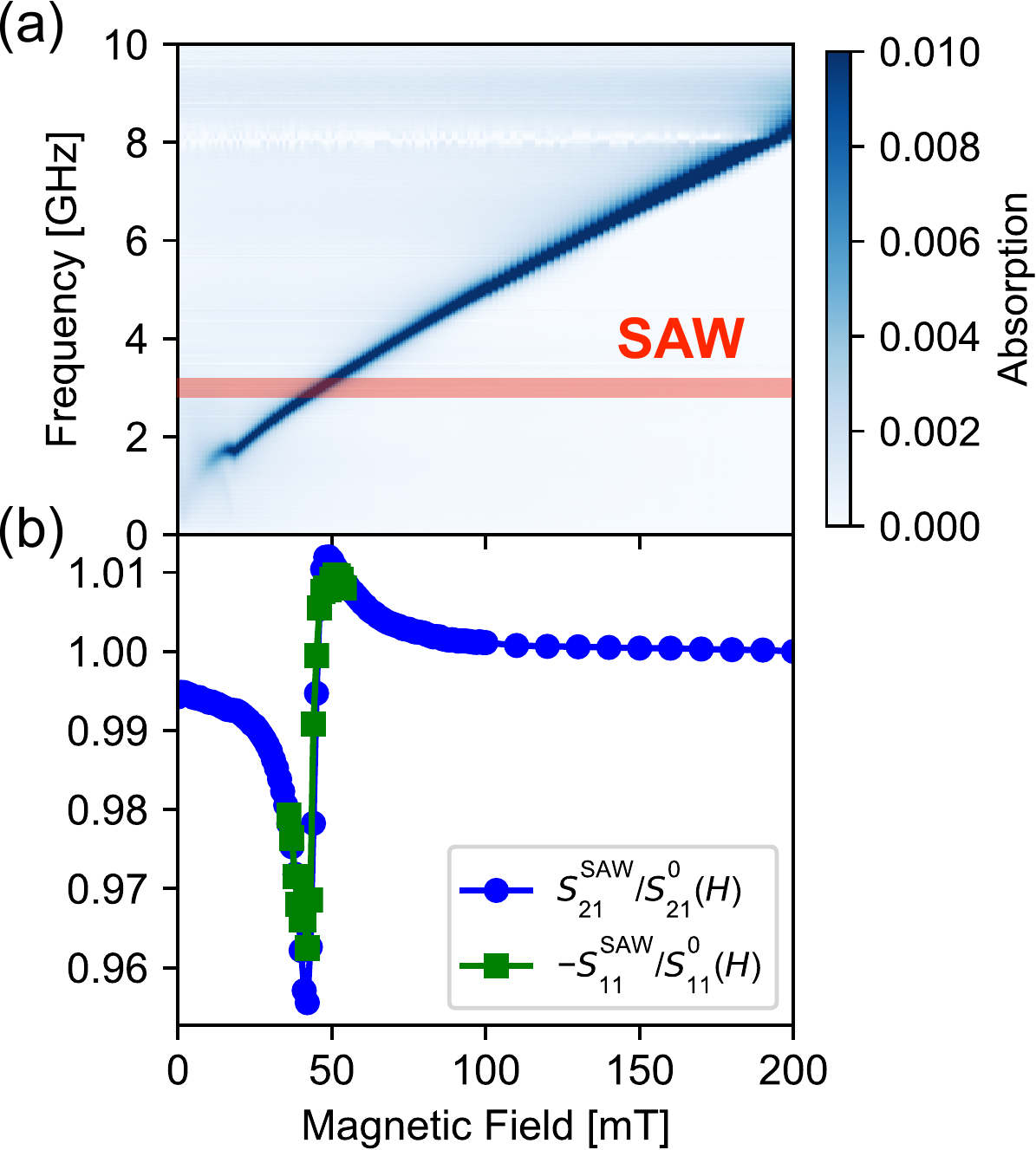}
  \caption{
  (a) Magnetic field dependence of microwave absorption due to magnetic resonance of CuB$_2$O$_4$. The external magnetic field is parallel to $\left[110\right]$ axis. Thick line shows the SAW frequency.
  (b) Magnetic field dependence of normalized SAW transmission $S_{21}^\textrm{SAW}/S^0_{21}$ and excitation $-S_{11}^\textrm{SAW}/S^0_{11}$.}
\end{figure}

Figure 2(a) shows the magnetic field dependence of microwave absorption spectra at $T$ = 10 K, which is measured in the measurement system similar to that of ref. 22 \cite{Nii2017}.
The static and alternating magnetic fields were parallel to $[110]$ axis and (110) plane, respectively. 
We have found a magnetic resonance mode in this frequency region. The frequency increases almost linearly with the magnetic field. As discussed in appendix, the origin is ascribed to the acoustic mode of antiferromagnetic resonance.
The magnetic resonance frequency coincided with the frequency of the SAW around 40 mT.

Figure 2(b) shows magnetic field dependence of  SAW transmission normalized by that at 200 mT, where the magnetic resonance frequency is far above the SAW frequency ($S^\textrm{SAW}_{21}/S_{21}^0(H)$; for the precise definition, see Appendix A)).
The SAW transmission gradually decreases with increasing magnetic field from 0 mT.
It shows rapid increase when the magnetic resonance frequency coincides with the SAW frequency.
Then it decreases toward the high field.
This characteristic magnetic field dependence seems caused by the interaction between the SAW and magnetic resonance.
In Fig. 2(b), we also plotted SAW excitation intensity at IDT 1 normalized by the 200 mT value ($-S_{11}/S_{11}^0(H)$; for the precise definition, see Appendix A)), which is estimated by the decrease of reflection due to SAW generation.
Because both the transmission and excitation show similar dependence on the magnetic field, it seems that the magnetic field dependence is caused by the excitation process.

\begin{figure}
  \centering
  \includegraphics[width = 12 cm]{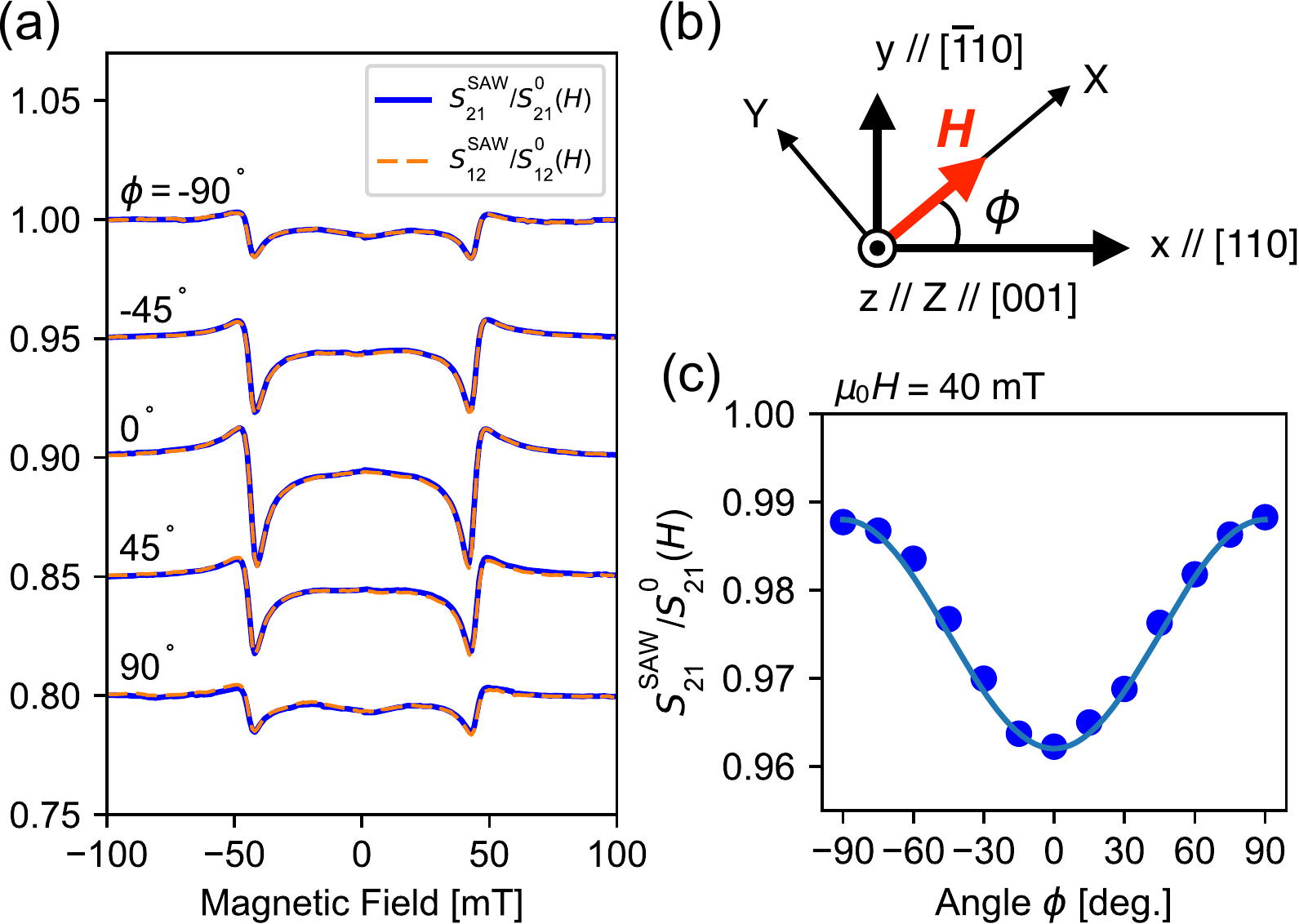}
  \caption{
  (a) Normalized SAW transmissions $S^{\rm SAW}_{21}$ and $S^{\rm SAW}_{12}$  as a function of magnetic fields with various directions. Solid lines show the transmission from port 1 to port 2 ($S^{\rm SAW}_{21}/S^0_{21}$) and  dashed lines are transmission for opposite direction ($S^{\rm SAW}_{12}/S^0_{12}$) .
  (b) Illustration of xyz- and XYZ-coordinate systems.
  (c) Magnetic field angle dependence of normalized SAW transmission $S^{\rm SAW}_{21}$ at $\mu_0 H$ = 40 mT. Solid line shows $C_1 - C_2\cos^2 \phi $.}
  \label{fig:3}
\end{figure}

In order to discuss the microscopic origin of magnetic field dependence, we show the SAW transmission as a function of the magnetic field along various directions $\phi$ in Fig.  3(a).
The magnetic field dependence became broad and the 
magnitude decreased with increasing or decreasing $\phi$ from 0$^\circ$.
Nevertheless, the characteristic magnetic field dependence around 40 mT was still discerned even at $\phi$ = $\pm$ 90$^\circ$.
Irrespective of the field angle, the magnetic field dependence is almost unchanged when the field direction or propagation direction is reversed ($S^\textrm{SAW}_{12}/S_{12}^0(H)$ is the normalized transmission for the opposite SAW propagation direction).
Figure 3(c) shows the magnetic field angle dependence of SAW transmission at 40 mT. The SAW transmission varied with $\phi$ as  $C_1-C_2\cos^2 \phi$ ($C_1$ and $C_2$ are constants).

Let us theoretically discuss the microscopic origin of the magnetic field dependence of the SAW signal.
When the frequency of magnetic resonance is close to that of SAW, the SAW excited state $\left|\textrm{SAW}\right>$ and magnetic excited state $\left|\textrm{Mag}\right>$  are expected to be hybridized with each other through the magnetoelastic coupling $H_{me}$. The SAW state hybridized with the magnetic excited state is expressed as
\begin{equation}
\left|\widetilde{\textrm{SAW}}\right>
\approx
\left|\textrm{SAW}\right>
+\frac{\left<\textrm{Mag}|H_{me}|\textrm{SAW}\right>}{\hbar\omega_{\textrm{Mag}} -\hbar\omega_{\textrm{SAW}}}
\left|\textrm{Mag}\right>,
\end{equation}
where $\omega_{\textrm{Mag}}$, $\omega_{\textrm{SAW}}$, and $\hbar$ are the frequencies of magnetic excitation and SAW, and Planck constant divided by $2\pi$, respectively.
When  $\omega_{\textrm{Mag}}$ is increased from 0, the magnitude of the second term in the right hand side increases. It shows steep sign change at $\omega_{\textrm{Mag}} = \omega_{\textrm{SAW}}$. Then the magnitude decreases toward 0 with  $\omega_{\textrm{Mag}}$. 
The magnetic field dependence of SAW excitation may be related to this.
The angle dependence is more closely related to the microscopic mechanism. To discuss this issue we should consider the explicit form of magnetoelastic coupling.
The magnetoelastic coupling energy in antiferromagnetic ordered  state is  
\begin{equation}
F_{me} = \sum_{p,q = 1,2}\sum_{i,j,k,l = 1,2,3}b_{pqijkl}m_{pi}m_{qj}\epsilon_{kl},
\end{equation}
where $m_{pi}$ represents the $i$-component of magnetization on $p$ sublattice of the antiferromagnet, and $\epsilon_{kl}$ is the strain tensor on the surface of the device.
The subscripts 1,2,3 for $i,j,k,l$ indicate $x,y,z$ components, respectively. 
The coefficients $b_{pqijkl}$ represent magnetoelastic coupling constants.
The nonvanishing component of $b_{pqijkl}$ is determined by the symmetry analysis (see appendix B).
The effective magnetic field $\bm{h}^{me}_p$ due to the strains acting on the magnetic moments at $p$-sublattice $\bm{m}_p$ is given by $\bm{h}^{me}_p = -\nabla_{\bm{m}_p}F_{me}  $.
To understand the acoustic antiferromagnetic resonance under SAW excitation, we consider  Landau-Lifshitz (LL) equation without damping term, in which the magnetic moments are driven by $\bm{h}^{me}_p$ as well as $\bm{H}$ and anisotropy field. By partially diagonalizing  6 components of the equation, we can deduce 3 effective equations regarding the acoustic antiferromagnetic resonance as follows (see appendix) ;
\begin{align}
\frac{1}{\gamma}\frac{\partial}{\partial t} \left( (\delta m_{1X} - \delta m_{2X}) + \frac{b}{a^2}(\delta m_{1Y}+\delta m_{2Y})\right) = \left(m^0\right)^2 \left(B_1 +\frac{b}{a^2} B_2 \right)\epsilon _{31}\cos \phi,\label{eq:ellipsoidal 0}\\
\left(\frac{1}{\gamma}\frac{\partial}{\partial t} -i a H^0\right)\left( (\delta m_{1Y} + \delta m_{2Y}) + i a(\delta m_{1Z}+\delta m_{2Z})\right) = \left(m^0\right)^2 \left(B_2\epsilon_{31}\cos \phi + i a B_3 \epsilon_{11}\sin 2 \phi\right), \label{eq:ellipsoidal 1}\\
\left(\frac{1}{\gamma}\frac{\partial}{\partial t} +i a H^0\right)\left( (\delta m_{1Y} + \delta m_{2Y}) - i a(\delta m_{1Z}+\delta m_{2Z})\right) = \left(m^0\right)^2 \left(B_2\epsilon_{31}\cos \phi - i aB_3 \epsilon_{11}\sin 2 \phi\right).\label{eq:ellipsoidal 2}
\end{align}
Here, $\gamma$ and $H^0$ are gyromagnetic ratio and magnitude of external magnetic field, respectively.
$\delta \bm{m_p}$ and $m^0$ are defined by $\delta \bm{m}_p = \bm{m}_p - \bm{m}_p^0$ and $m^0 = |\bm{m}^0_p|$, where $\bm{m}^0_p$ is static part of magnetic moments. $XYZ$ coordinate system is defined as in Fig.  3(b), so that $X$-axis is parallel to the magnetic field. Dimensionless constants $a$ and $b$ are defined by $a = \sqrt{\left(-\frac{K}{m^0}\cos\psi + H^0\right)/H^0}$ and $b = \left(-\frac{K}{m^0}+2m^0\Lambda\right)\sin\psi /H^0$, where $K,\Lambda$ and $\psi$ are the uniaxial magnetic anisotropy constant, molecular field constant, and angle between the sublattice magnetic moments and magnetic field, respectively.
$B_1$,$B_2$ and $B_3$ are constants defined by
\begin{align}
B_1 = 8 \sin \psi \left(\left(b_{111313}+b_{121313}\right) \cos \psi-b_{121323} \sin \psi\right),\\
B_2 = -8 \cos \psi \left(\left(b_{111313}+b_{121313}\right) \cos \psi-b_{121323} \sin \psi\right),\\
B_3 = -2 \left(\left(b_{111111}-b_{111122}\right) \cos 2 \psi+b_{121111}-b_{121122}\right).
\end{align}
In the absence of SAW excitation, the right hand sides of Eq.(\ref{eq:ellipsoidal 0})-(\ref{eq:ellipsoidal 2}) vanish. In this case, these equations stand for the pure antiferromagnetic excitation. 
Because $\omega_{\textrm{Mag}} > 0$, only right-hand ellipsoidal polarized components $ (\delta m_{1Y} + \delta m_{2Y}) + i a(\delta m_{1Z}+\delta m_{2Z})$ exhibits resonance behavior at $\omega_{\textrm{Mag}} = \gamma a H^0$. Eq. (\ref{eq:ellipsoidal 2}) does not show any resonance behavior, and Eq. (\ref{eq:ellipsoidal 0}) stand for constraint condition describing the relationship between magnetic moment components parallel ($\delta m_{1X}, \delta m_{2X}$) and perpendicular ($\delta m_{1Y}, \delta m_{2Y}$) to the magnetic field during the precession motion of magnetic moments. The right hand sides of these equations are the magnetic torque due to the SAW excitation. In particular, the right hand side of Eq. (\ref{eq:ellipsoidal 1}) is the direct coupling between the SAW and magnetic resonance mode. The magnitude of hybridization  is proportional to 
\begin{equation}
\left|B_2\epsilon_{31}\cos \phi + i a B_3 \epsilon_{11}\sin 2 \phi\right|^2.
\end{equation}

The experimentally obtained magnetic field dependence of   $C_1 -C_2\cos^2\phi$ indicates  
that $B_3$ is negligible, which is composed of longitudinal type magnetoelastic coupling constants.  It seems that the strain $\epsilon_{31}$ most effectively excite the acoustic antiferromagnetic resonance. 
Note that nonreciprocity should be caused by the mixture of phase different effective magnetic fields induced by the shear-type and longitudinal acoustic strains, similarly to the case of Ni/LiNbO$_3$\cite{Sasaki2017}.  In this case, however, the longitudinal type magnetoelastic coupling is absent, and the magnetic field is linearly polarized. That is why the nonreciprocity is negligible in this system. 
The constant term $C_1$ cannot be deduced by the present theoretical analysis, in which the magnitude of magnetic moments is fixed. Perhaps, the constant term is caused by the magnetic field change of the magnitude.

In conclusion, we could successfully excite and detect the surface acoustic wave on multiferroic material CuB$_2$O$_4$ by fabricating the interdigital transducers on a  CuB$_2$O$_4$ single crystal substrate.
In the antiferromagnetic phase, SAW excitation and transmission exhibited characteristic change due to the coupling to the antiferromagnetic resonance.  This magnetic field dependence was explained by the analysis using the effective magnetic field caused by magnetoelastic coupling. 
This research may pave a new path to increase the controllability of the SAW device.

\begin{acknowledgments}
This work was in part supported by the
Grant-in-Aid for Scientific Research (Grants No. 17H05176,
No. 16H04008) and for Young Scientists (18K13494) from the Japan Society for the Promotion
of Science and the Noguchi Institute. R.S. is supported by the Grant-in-Aid for JSPS
Research Fellow (No. 18J12130).
\end{acknowledgments}

\appendix
\setcounter{figure}{0} \renewcommand{\thefigure}{A.\arabic{figure}}
\section{Definitions of relative SAW excitation and transmission $-S^\textrm{SAW}_{11}/S^0_{11}(H)$ and $S^\textrm{SAW}_{21}/S^0_{21}(H)$}
\begin{figure}
  \centering
  \includegraphics[width = 12 cm]{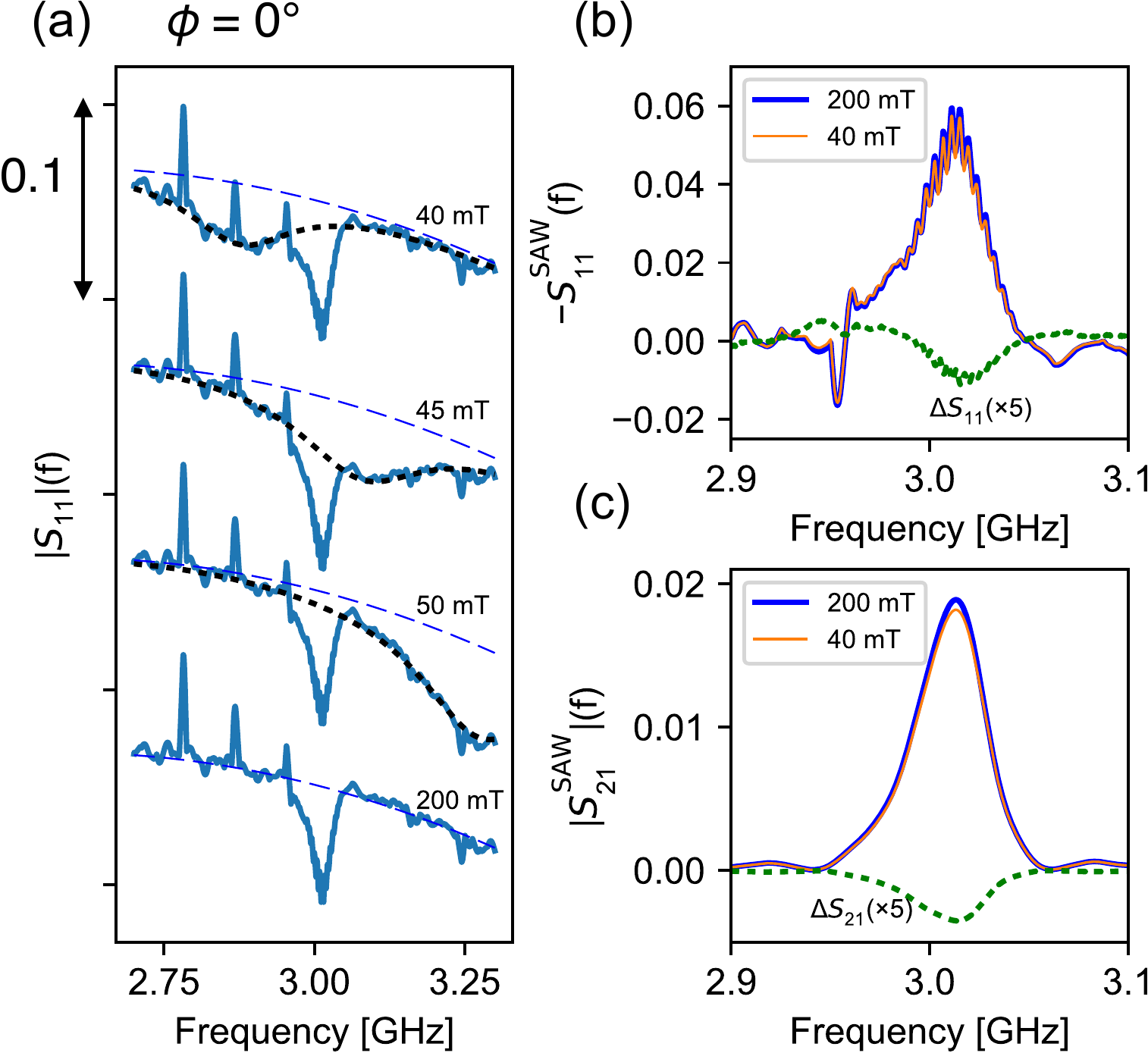}
  \caption{
   (a) Microwave reflection  from the SAW device $|S_{11}|(f)$ at several magnetic fields. The magnetic fields are applied parallel to the SAW propagation direction. The solid line shows the observed data, and the dotted and dashed lines are the result of fitting of magnetic resonance and background, respectively (see text).
  (b) SAW excitation $-S^{\rm SAW}_{11}$ at 40 mT and 200 mT and the difference between 40 mT and 200 mT values ($\Delta S_{11}$). 
  (c) SAW transmission $S^{\rm SAW}_{21}$ at 40 mT and 200 mT and their difference ($\Delta S_{21}$). $\Delta S_{11}$ and $\Delta S_{21}$ in (b) and (c) are multiplied by 5.}
  \label{fig:app1}
\end{figure}
Figure \ref{fig:app1}(a) shows reflection  of the device $|S_{11}|(f)$ under the external magnetic fields $\mu_0 H$ = 40, 45, 50, and 200 mT. The external magnetic field is applied parallel to the SAW propagation direction, i.e. $\phi = 0^\circ$.
There are three components of decrease of reflection  (absorption); a sharp dip, broad dip, and background signal from the device. While  the sharp peak and background do not show large magentic field variation, the frequency of broad peak increases with the magnetic field. Therefore, we ascribed the origins of sharp and broad peaks to SAW and magnetic resonance, respectively.
In order to accurately estimate the SAW contribution, we fitted the contribution of magnetic resonance and background to the Lorentzian and quadratic function, respectively, and subtract them from $|S_{11}|(f)$.
Figure \ref{fig:app1}(b) shows the minus subtracted reflection $-S_{11}^{\rm SAW}(f)$ at $\mu_0 H$ = 40 and 200 mT. 
Because the frequency of magnetic resonance is far above 3 GHz at 200 mT, the SAW excitation is not affected by the magnetic resonance. When the magnetic field is decreased to 40 mT, the SAW intensity decreases (Fig. A.1.(b)). Similarly, the transmission at 40 mT is a little smaller than that  at 200 mT as shown in Fig. A.1 (c). The relative  change of SAW reflection and transmission is defined as 
\begin{align}
S^\textrm{SAW}_{11}/S_{11}^0(H) &= {\rm max}\left\{|S^{\rm SAW}_{11}|(f)\right\}(H)/{\rm max}\left\{|S^{\rm SAW}_{11}|(f)\right\}(H = 200\ \rm mT/\mu_0 ),\\
S^\textrm{SAW}_{21}/S_{21}^0(H) &= {\rm max}\left\{|S^{\rm SAW}_{21}|(f)\right\}(H)/{\rm max}\left\{|S^{\rm SAW}_{21}|(f)\right\}(H = 200\ \rm mT/\mu_0 )
\end{align}
where max$\{\}$ stands for the maximum value.
From the definition of SAW reflection, the value $-S^\textrm{SAW}_{11}/S_{11}^0(H)$ can be interpreted as SAW excitation.
These quantities are plotted in Fig. 2(b) in the main text.

\section{magnetoelastic conpling energy}
As discussed above, the magnetoelastic coupling energy in staggered antiferromagnetic state can be expressed as
\begin{equation}
F_{me} = \sum_{p,q = 1,2}\sum_{i,j,k,l = 1,2,3}b_{pqijkl}m_{pi}m_{qj}\epsilon_{kl}.
\end{equation}
Because the magnetoelastic coupling energy is unchanged by the symmetry operation belong to the space group $I\bar{4}2d$, some of constants $b_{pqijkl}$ should vanish and the number of independent magnetoelastic coupling constants decrease. The reduced magnetoelastic coupling energy can be represented by the following equation;
\begin{equation}\label{eq: reduced me energy}
\begin{aligned}
F_{me}
&=\epsilon _{11}(%
b_{111111} m_{11}^2+2 b_{121111} m_{21} m_{11}+b_{111111} m_{21}^2+b_{111122} \left(m_{12}^2+m_{22}^2\right)\\
&+b_{113311} \left(m_{13}^2+m_{23}^2\right)\\
&+2 b_{121122} m_{12} m_{22}+2 b_{121211} \left(m_{11} m_{22}-m_{12} m_{21}\right)+2 b_{123311} m_{13} m_{23}
)\\
&+\epsilon _{12}(%
4 b_{111212} \left(m_{11} m_{12}+m_{21} m_{22}\right)+4 b_{121212} \left(m_{12} m_{21}+m_{11} m_{22}\right)
)\\
&+\epsilon _{13}(%
4 (m_{21} \left(b_{111313} m_{23}+b_{121313} m_{13}\right)+m_{11} \left(b_{111313} m_{13}+b_{121313} m_{23}\right)\\
&+b_{121323} \left(m_{13} m_{22}-m_{12} m_{23}\right))
)\\
&+\epsilon _{22}(%
b_{111111} m_{12}^2+2 b_{121111} m_{22} m_{12}+b_{111111} m_{22}^2+b_{111122} \left(m_{11}^2+m_{21}^2\right)\\
&+b_{113311} \left(m_{13}^2+m_{23}^2\right)+2 b_{121122} m_{11} m_{21}+2 b_{121211} \left(m_{11} m_{22}-m_{12} m_{21}\right)+2 b_{123311} m_{13} m_{23}
)\\
&+\epsilon _{23}(%
4(m_{22} \left(b_{111313} m_{23}+b_{121313} m_{13}\right)+m_{12} \left(b_{111313} m_{13}+b_{121313} m_{23}\right)\\
&+b_{121323} \left(m_{11} m_{23}-m_{13} m_{21}\right))
)\\
&+\epsilon _{33}(%
b_{111133} m_{11}^2+2 m_{11} \left(b_{121133} m_{21}+b_{121233} m_{22}\right)+b_{111133} m_{12}^2+b_{111133} \left(m_{21}^2+m_{22}^2\right)\\
&+b_{113333} \left(m_{13}^2+m_{23}^2\right)+2 m_{12} \left(b_{121133} m_{22}-b_{121233} m_{21}\right)+2 b_{123333} m_{13} m_{23}
).
\end{aligned}
\end{equation}

\section{Effective equations of motion for the antiferromagnetic resonances under the SAW excitations}

The effective magnetic field on the magnetization $\bm{m}_p$ ($p$ = 1,2) is defined by 
\begin{equation}
\bm{h}^{me}_p = -\bm{\nabla}_{\bm{m}_p}F_{me} =(h^{me}_{p1},h^{me}_{p2},h^{me}_{p3}).
\end{equation}
The subscript $p$ represents the index of two sublattice of ordered spins.

Assuming that $\bm{h}^{me}_p$ acts on the magnetic moments similarly to the real magnetic field , LL equation without damping term is represented by 
\begin{equation}
\frac{\partial \bm{m}_p}{\partial t} = -\gamma \bm{m}_p\times \left\{ -\Lambda \bm{m}_q + \frac{K}{|\bm{m}_p|^2}(\bm{m}_p\cdot \hat{\bm{z}})\hat{\bm{z}} + \bm{h}^{me}_p +\bm{H}\right\}
,
\end{equation}
where $\hat{\bm{}{z}}$ is the unit vector along $z$-axis.  By putting $\bm{m}_p=\bm{m}_p^0+\delta \bm{m}_p$ and  linearizing the 6 equations with respect to $\delta\bm{m}$ and $\bm{h}^{me}_p$, we get
\begin{equation}\label{eq app:linearized LL equation}
\frac{1}{\gamma}\frac{\partial \delta\bm{m}_p}{\partial t} +   \bm{m}_p^0 \times(-\Lambda\delta\bm{m}_q + \frac{K}{(m^0)^2}(\delta\bm{m}_p\cdot \hat{\bm{z}})\hat{\bm{z}} )  + \delta \bm{m}_p\times(-\Lambda\bm{m}_q^0 + \bm{H}) =- \bm{m}_p^0 \times \bm{h}^{me}_p
,
\end{equation}
where $p,q$ = 1 or 2 and $p \neq q$. 

Based on the reduced magnetoelastic coupling energy (\ref{eq: reduced me energy}) , the effective magnetic field components form
\begin{equation}
\begin{split}
h^{me}_{11} =& -2 m^0_{11} \left(b_{111111} \epsilon _{11}+b_{111122} \epsilon _{22}+b_{111133} \epsilon _{33}\right)\\
&-4 b_{111212} m^0_{12} \epsilon _{12}-2 m^0_{21} \left(b_{121111} \epsilon _{11}+b_{121122} \epsilon _{22}+b_{121133} \epsilon _{33}\right)\\
&-2 m^0_{22} \left(b_{121211} \left(\epsilon _{11}+\epsilon _{22}\right)+2 b_{121212} \epsilon _{12}+b_{121233} \epsilon_{33}\right),\\
h^{me}_{12} =& -2 b_{111111} m^0_{12} \epsilon _{22}-2 b_{111122} m^0_{12} \epsilon _{11}-2 b_{111133} m^0_{12} \epsilon _{33}-4 b_{111212} m^0_{11} \epsilon _{12}-2 b_{121111} m^0_{22} \epsilon _{22}\\
&-2 b_{121122} m^0_{22} \epsilon _{11}-2 b_{121133} m^0_{22} \epsilon _{33}+2 m^0_{21} \left(b_{121211} \left(\epsilon _{11}+\epsilon _{22}\right)-2 b_{121212} \epsilon _{12}+b_{121233} \epsilon _{33}\right),\\
h^{me}_{13} =& -4 b_{111313} m^0_{11} \epsilon _{31}-4 b_{111313} m^0_{12} \epsilon _{23}-4 b_{121313} m^0_{22} \epsilon _{23}-4 b_{121323} m^0_{22} \epsilon _{31}\\
&+2 m^0_{21} \left(2 b_{121323} \epsilon _{23}-2 b_{121313} \epsilon _{31}\right),\\
h^{me}_{21} =& -2 m^0_{21} \left(b_{111111} \epsilon _{11}+b_{111122} \epsilon _{22}+b_{111133} \epsilon _{33}\right)-2 m^0_{11} \left(b_{121111} \epsilon _{11}+b_{121122} \epsilon _{22}+b_{121133} \epsilon _{33}\right)\\
&+2 \left( m^0_{12} \left(b_{121211} \left(\epsilon _{11}+\epsilon _{22}\right)-2 b_{121212} \epsilon _{12}+b_{121233} \epsilon _{33}\right)-2 b_{111212} m^0_{22} \epsilon _{12}\right),\\
h^{me}_{22} =& -2 b_{111111} m^0_{22} \epsilon _{22}-2 b_{111122} m^0_{22} \epsilon _{11}-2 b_{111133} m^0_{22} \epsilon _{33}-4 b_{111212} m^0_{21} \epsilon _{12}-2 b_{121111} m^0_{12} \epsilon _{22}\\
&-2 b_{121122} m^0_{12} \epsilon _{11}-2 b_{121133} m^0_{12} \epsilon _{33}-2 m^0_{11} \left(b_{121211} \left(\epsilon _{11}+\epsilon _{22}\right)+2 b_{121212} \epsilon _{12}+b_{121233} \epsilon _{33}\right),\\
h^{me}_{23} =& -4 b_{111313} m^0_{21} \epsilon _{31}-4 b_{111313} m^0_{22} \epsilon _{23}-4 b_{121313} m^0_{12} \epsilon _{23}+4 b_{121323} m^0_{12} \epsilon _{31}\\&
-4 m^0_{11} \left(b_{121313} \epsilon _{31}+b_{121323} \epsilon _{23}\right).
\end{split}
\end{equation}
Here we estimate the effective magnetic field due to the small strain. Therefore, we neglect the dynamical component of magnetization ($\delta \bm{m}_p$).
Since the magnetic order in CuB$_2$O$_4$ at $T$= 10 K is easy-plane N\'eel type magnetic order, we assume $m_{p3}^0=0$. 

From the fact that the device was designed so that its sagittal plane is parallel to the mirror plane of the crystal, we can assume that Rayleigh type SAW mode is excited in our measurement\cite{mason2012physical}. 
We  neglected the strain components other than $\epsilon_{11}$, $\epsilon_{33}$, and $\epsilon_{31}$, of which Rayleigh type SAW is composed.
By diagonalizing Eq. (\ref{eq app:linearized LL equation}), we get the six equations for antiferromagnetic resonance as follows;
\begin{align}
\frac{1}{\gamma}\frac{\partial}{\partial t} \left( (\delta m_{1X} - \delta m_{2X}) + \frac{b}{a^2}(\delta m_{1Y}+\delta m_{2Y})\right) = \left(m^0\right)^2 \left(B_1 +\frac{b}{a^2} B_2 \right)\epsilon _{31}\cos \phi,\label{eq app:ellipsoidal 0}\\
\left(\frac{1}{\gamma}\frac{\partial}{\partial t} -i a H^0\right)\left( (\delta m_{1Y} + \delta m_{2Y}) + i a(\delta m_{1Z}+\delta m_{2Z})\right) = \left(m^0\right)^2 \left(B_2\epsilon_{31}\cos \phi + i a B_3 \epsilon_{11}\sin 2 \phi\right), \label{eq app:ellipsoidal 1}\\
\left(\frac{1}{\gamma}\frac{\partial}{\partial t} +i a H^0\right)\left( (\delta m_{1Y} + \delta m_{2Y}) - i a(\delta m_{1Z}+\delta m_{2Z})\right) =\left(m^0\right)^2 \left(B_2\epsilon_{31}\cos \phi - i aB_3 \epsilon_{11}\sin 2 \phi\right),\label{eq app:ellipsoidal 2}\\
\frac{1}{\gamma}\frac{\partial}{\partial t} \left( (\delta m_{1X} + \delta m_{2X}) + c(\delta m_{1Y} - \delta m_{2Y})\right) = \left(m^0\right)^2 \left(B_4 +c B_5\right)\epsilon_{13}\cos \phi \label{eq app: optical 0},\\
\left(\frac{1}{\gamma}\frac{\partial}{\partial t} -i e(2m\Lambda\cos\psi- H^0)\right)\left( d(\delta m_{1X} + \delta m_{2X}) + (\delta m_{1Y} - \delta m_{2Y}) - i e(\delta m_{1Z}-\delta m_{2Z})\right) \nonumber\\
=  \left(m^0\right)^2 \left(dB_4 + B_5\right)\epsilon_{13}\cos \phi -ie(B_6 \epsilon_{33}+\epsilon_{11}(B_7\cos 2 \phi  +B_8))\label{eq app: optical 1},\\
\left(\frac{1}{\gamma}\frac{\partial}{\partial t} +i e(2m\Lambda\cos\psi- H^0)\right)\left( d(\delta m_{1X} + \delta m_{2X}) + (\delta m_{1Y} - \delta m_{2Y}) + i e(\delta m_{1Z}-\delta m_{2Z})\right) \nonumber\\
=  \left(m^0\right)^2 \left(dB_4 + B_5\right)\epsilon_{13}\cos \phi + ie(B_6 \epsilon_{33}+\epsilon_{11}(B_7\cos 2 \phi  +B_8))\label{eq app: optical 2},
\end{align}
where $\delta m_{1X} = \delta m_{11}\cos \phi +\delta m_{12}\sin \phi ,
\delta m_{2X} = \delta m_{21}\cos \phi +\delta m_{22}\sin \phi ,
\delta m_{1Y} = -\delta m_{11}\sin \phi +\delta m_{12}\cos \phi ,
\delta m_{2Y} = -\delta m_{21}\sin \phi +\delta m_{22}\cos \phi ,
\delta m_{1Z} = \delta m_{13},
\delta m_{2Z} = \delta m_{23}.
$
Dimensionless constants $c,d$, and $e$ are  defined by
$c =  \frac{K \sin \psi}{\left(2 \Lambda  (m^0)^2+K\right) \cos \psi-m^0 H^0}$,
$d =   \frac{2 m^0 \Lambda  \sin \psi}{2 m^0 \Lambda  \cos \psi - H^0}$, 
$e= \frac{\sqrt{\left(H^0\right)^2-\left(4 \Lambda  (m^0)+K/m^0\right) \cos \psi H^0+2 \Lambda  \left(\Lambda  (m^0)^2+\left(\Lambda  (m^0)^2+K\right) \cos 2 \psi\right)}}{ 2 m^0 \Lambda  \cos \psi-H^0} $.
$B_4,B_5$ and $B_6$ are constants defined by
\begin{align}
B_4 = 8 \sin \psi \left(\left(b_{121313}-b_{111313}\right) \sin \psi+b_{121323} \cos \psi\right),\\
B_5 = -8 \cos \psi \left(\left(b_{121313}-b_{111313}\right) \sin \psi+b_{121323} \cos \psi\right),\\
B_6 = -4 \left(b_{121133} \sin 2 \psi+b_{121233} \cos 2 \psi\right),\\
B_7 = -2 \sin 2 \psi \left(b_{111111}-b_{111122}\right) ,\\
B_8 = -2 \left(\left(b_{121111}+b_{121122}\right) \sin 2 \psi+2 b_{121211} \cos 2 \psi\right).
\end{align} 
Eqs.(\ref{eq app:ellipsoidal 0})-(\ref{eq app:ellipsoidal 2}) and Eqs.(\ref{eq app: optical 0})-(\ref{eq app: optical 2}) are relevant to the acoustic mode and optical mode of antiferromagnetic resonance, respectively.

\section{Antiferromagnetic resonance measurements}

\begin{figure}
  \centering
  \includegraphics[width = 8 cm]{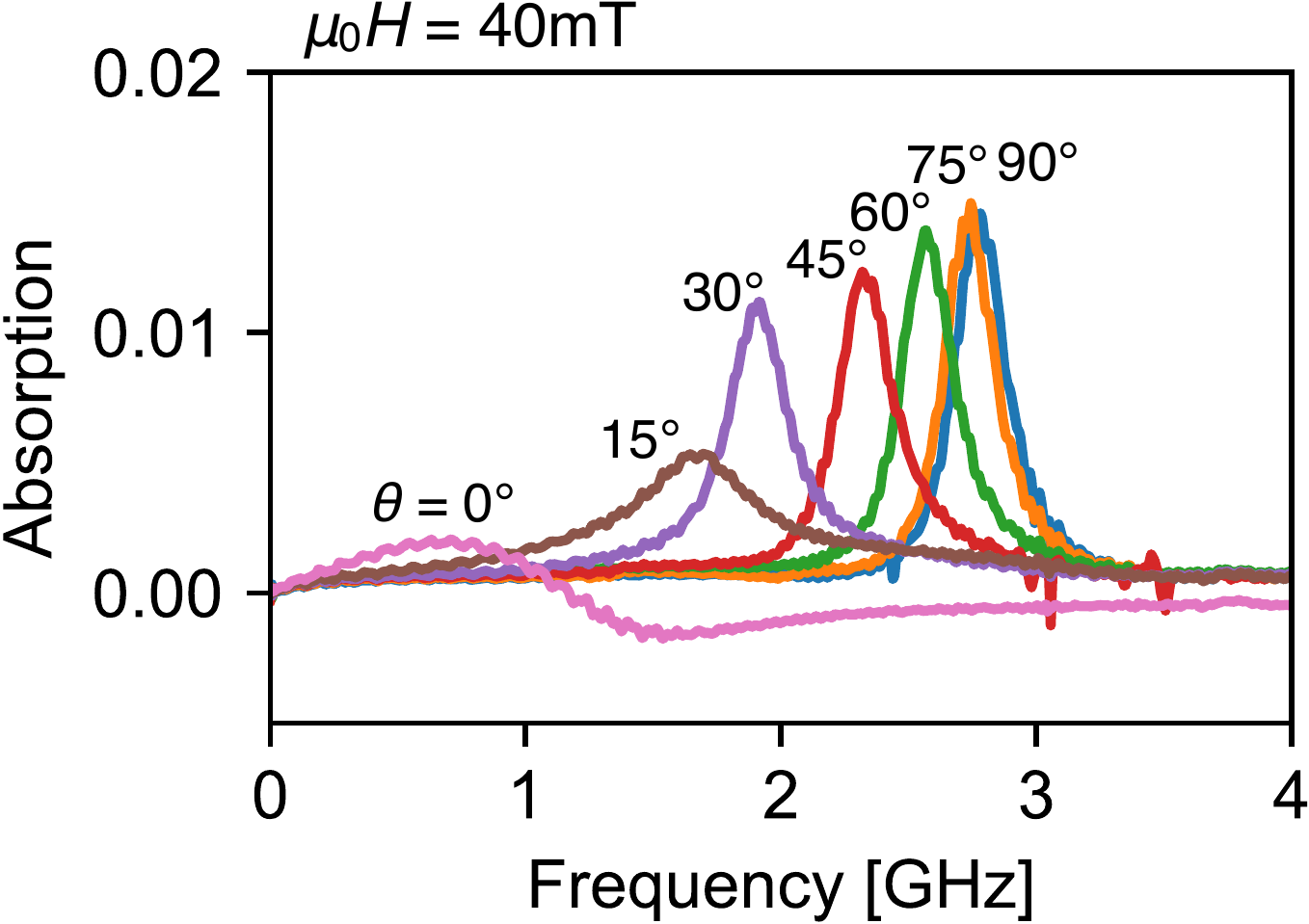}
  \caption{
  Microwave absorption spectra in magnetic fields with various directions within the ($1\bar{1}0$) plane and the fixed magnitude of $\mu_0H$ = 40 mT. $\theta$ is defined as the angle between the $[001]$ axis and the magnetic field. 
  }
  \label{fig:app2}
\end{figure}

We measured the microwave absorption of CuB$_2$O$_4$ in order to study the origin of magnetic resonance. While the magnetic field is along the (001) plane in the previous work\cite{Nii2017}, it is within the ($1\bar{1}0$) plane in the present study.  The other experimental configurations are similar to those in the previous study. 
Figure \ref{fig:app2} shows microwave absorption in magnetic fields with various directions and the fixed magnitude of 40 mT.
$\theta$ is defined as the angle between the [001] axis and the magnetic field.
As the angle $\theta$ decrease, magnetic resonance spectra shifted to low frequency and its intensity decreases. This is quit consistent with the picture of magnetic resonance in easy plane antiferromagnets because the lowest magnetic resonance becomes zero mode in the plane-normal magnetic field for the easy plane antiferromagnets\cite{gurevich1996magnetization}. In the previous work, we tentatively assigned this magnetic resonance to the paramagnetic resonance of Cu(B) site because this absorption peak is smoothly connected to the magnetic resonance peak above $T_N$. Nevertheless, this new experimental result strongly supports the antiferromagnetic resonance origin.

\nocite{*}
\bibliography{aipsamp}

\end{document}